\documentstyle[aps]{revtex}

\begin{document}
\title{Branching ratio for $B\rightarrow K_{1}\gamma $ decay in next-to-leading
order in LEET}
\author{M. Jamil\ Aslam}
\address{National Centre for Physics and Department of Physics,\\
Quaid-i-Azam University, \\
Islamabad, Pakistan}
\author{Riazuddin}
\address{National Centre for Physics,\\
Quaid-i-Azam University, \\
Islamabad, Pakistan}
\maketitle

\begin{abstract}
Branching ratio for $B\rightarrow K_{1}\gamma $ at next-to-leading order of $%
\alpha _{s}$ has been calculated in Large Energy Effective Theory. By
incorporating the higher twist effects in light cone decay amplitude for
axial $K$ meson, it is shown that the form factor is not sensitive to these
twists.
\end{abstract}

\section{Introduction}

Rare $B$ decays involving flavor-changing-neutral-current (FCNC)
transitions, such as $b\rightarrow s\gamma $, have received a lot of
theoretical interest \cite{Greub:1999sv}. First measurements of the decay $%
B\rightarrow X_{s}\gamma $ were reported by the CLEO collaboration \cite
{Alam:1995aw}. These decays are now being investigated more precisely in
experiments at the B factories. The current world average based on the
improved measurements by the CLEO \cite{Chen:2001fj}, ALEPH \cite{alephbsg}
and BELLE collaborations, ${\cal B}(B\rightarrow X_{s}\gamma )=(3.22\pm
0.40)\times 10^{-4}$, is in good agreement with the estimates of the
standard model (SM) \cite{Chetyrkin:1997vx,Kagan:1999ym,Gambino:2001ew},
which we shall take as ${\cal B}(B\rightarrow X_{s}\gamma )=(3.50\pm
0.50)\times 10^{-4}$, reflecting the parametric uncertainties dominated by
the scheme-dependence of the quark masses. The decay $B\rightarrow
X_{s}\gamma $ also provides useful constraints on the parameters of the
supersymmetric theories, which in the context of the minimal supersymmetric
standard model (MSSM) have been recently updated\cite{Ali:2001jg}.

Exclusive decays involving the $b\rightarrow s\gamma $ transition are best
exemplified by the decay $B\rightarrow K^{*}\gamma $, which provide abundant
issues for both theorists and experimentalists. After the first measurement
at CLEO, $B\rightarrow K^{*}\gamma $ is now also measured in Belle and
Babar: 
\begin{eqnarray}
{\cal B}(B^{0} &\rightarrow &K^{*0}\gamma )=\left\{ 
\begin{array}{c}
(4.09\pm 0.21\pm 0.19)\times 10^{-5}\,\,\,\,\,\,\text{Belle\cite{Belleee}}
\\ 
(4.23\pm 0.40\pm 0.22)\times 10^{-5}\,\,\,\,\,\text{BaBar\cite{BaBar}} \\ 
(4.55\pm 0.70\pm 0.34)\times 10^{-5}\,\,\,\,\,\text{CLEO\cite{CLEO}}
\end{array}
\right.  \label{data1} \\
{\cal B}(B^{+} &\rightarrow &K^{*+}\gamma )=\left\{ 
\begin{array}{c}
(4.40\pm 0.33\pm 0.24)\times 10^{-5}\,\,\,\,\,\,\text{Belle\cite{Belleee}}
\\ 
(3.83\pm 0.62\pm 0.22)\times 10^{-5}\,\,\,\,\,\text{BaBar\cite{BaBar}} \\ 
(3.76\pm 0.86\pm 0.28)\times 10^{-5}\,\,\,\,\,\text{CLEO\cite{CLEO}}
\end{array}
\right.  \label{data2}
\end{eqnarray}
On theoretical side there have been noticeable advances in $B\rightarrow
K^{*}\gamma $ for a decade. QCD corrections at next-to-leading order (NLO)
of $O\left( \alpha _{s}\right) $ have already been considered\cite
{Soares,Greub,Hurth}. Relevant Wilson coefficients have been improved up to
three loop level calculations\cite{Adel,Chetyrkin}. Recent developments of
the QCD factorizations helped one to calculate the hard spectator
contributions systematically in the factorized form through the convolution
at the heavy quark limit\cite{Feldmann,Siedel,Bosch}. The detailed analysis
of $B\rightarrow K^{*}\gamma $ has also been done at next to leading order
in effective theories, such as large energy effective theory (LEET) \cite
{Ali}, and in soft-collinear effective theory (SCET)\cite{Chay}.

In addition to $K^{*}$, higher resonances of kaon also deserve much
attention. Recently, Belle has announced the first measurement of $%
B\rightarrow K_{1}^{+}(1270)\gamma $ \cite{belle4led} 
\begin{equation}
{\cal{ B}} (B^+ \rightarrow K_1^+ \gamma ) = (4.28\pm 0.94\pm 0.43)\times
10^{-5}  \label{01}
\end{equation}
Among many reasons to focus on the higher resonances, the first one is that
they share lot of the things with $B\rightarrow K^{*}\gamma $, like at quark
level both of them are governed by $b\rightarrow s\gamma $. Therefore all
the achievements of $b\rightarrow s\gamma $ can be used in these decays,
e.g. the same operators in the operator product expansion and the same
Wilson coefficients that are available. The light cone distribution
amplitudes (DA) are same except the overall factor of $\gamma _{5}$ and this
gives few differences in many calculations\cite{Lee}. Secondly, it was
suggested that $B\rightarrow K_{\text{res}}\left( \rightarrow K\pi \pi
\right) \gamma $ can provide a direct measurement of the photon polarization%
\cite{Gronau} and it was shown that large polarization asymmetry $\approx
33\%$ has been produced due to decay of $B$ meson through the kaon
resonances. In the presences of anomalous right-handed couplings, the
polarization can be severely reduced in the parameter space allowed by
current experimental bounds of $B\rightarrow X_{s}\gamma $. It was also
argued that the $B$ factories can now make a lot of $B\bar{B}$ pairs, enough
to check the anomalous couplings through the measurement of the photon
polarization.

The theorists are also facing challenges from the discrepancy between their
predictions and experiments. It was pointed out that the form factor
obtained using the LEET approach for $B\rightarrow K^{*}\gamma $ is found to
be smaller compared to the values obtained by QCD sum rules or light-cone
sum rules (LCSR)\cite{Ali}. At this stage, the source of this mismatch is
not well understood.

On $B\rightarrow K_{1}\gamma $ side the situation is more complicated. Based
on the QCDF framework combined with the LCSR results, it is predicted that $%
{\cal B}(B^{0}\rightarrow K_{1}^{0}(1270)\gamma )=(0.828\pm 0.335)\times
10^{-5}$ at the NLO of $\alpha _{s}$ which is very small as compared to the
experimental value [cf. Eq. (\ref{01})] \cite{Lee}. The value of the
relevant form factor has been extracted from the experimental data and its
value is found to be $F_{+}^{K_{1}(1270)}(0)=0.32\pm 0.03$ which is very
large as compared to $F_{+}^{K_{1}(1270)}(0)|_{{\rm LCSR}}=0.14\pm 0.03$
obtained by the LCSR. These are contrary to the case of $B\rightarrow
K^{*}\gamma $ where the form factor obtained from LCSR is larger than the
LEET\ one and the source of discrepancy is not yet known. But for $%
B\rightarrow K_{1}\gamma $ case the possible candidates to explain this
discrepancy have also been discussed in detail in the literature\cite{Leenew}%
.

In this paper the branching ratio for $B\rightarrow K_{1}\gamma $ at NLO of $%
\alpha _{s}$ are calculated using the LEET\ approach\cite{Dugan,Charles}. We
follow the same frame work as done by Ali et. al.\cite{Ali} for $%
B\rightarrow K^{*}\gamma $, because $B\rightarrow K_{1}\gamma $ shares many
things with it. The only difference is the DA for the daughter meson. As $%
K_{1}$ is an axial vector and is distinguished by vector by the $\gamma _{5}$
in the gamma structure of DA and some non perturbative parameters. But the
presence of $\gamma _{5}$ does not alter the calculation, give the same
result for the perturbative part. The higher twist terms are also included
through the Gegenbauer moments in the Gegenbauer expansion. The calculation
with out Gegenbauer has already been done in QCD factorization frame work
and using the LCSR results for form factors and decay constant\cite
{Lee,Leenew}.

At next-to-leading order of $\alpha _{s}$ there are the contributions from
the operators $O_{2}$ $O_{7}$ and $O_{8}$ which will be discussed in detail.
The paper is organized as follows. In Section II we give the short flavor of
the leading order calculation for $B\rightarrow K_{1}\gamma $ decay process.
The section III deals with the hard spectator contributions in $B\rightarrow
K_{1}\gamma $ decays while in section IV $O\left( \alpha _{s}\right) $
corrected matrix element for above mentioned decays is discussed in detail.
The resulting branching ratio and related discussion appear in sec. V. The
concluding remarks are given at the end.

\section{Leading order contributions}

The effective Hamiltonian for $b\rightarrow s\gamma $ can be written as 
\begin{equation}
{\cal H}_{{\rm eff}}(b\rightarrow s\gamma )=-\frac{G_{F}}{\sqrt{2}}%
V_{tb}V_{ts}^{*}\sum_{i=1}^{8}C_{i}(\mu )O_{i}(\mu )~,  \label{02}
\end{equation}
where 
\begin{eqnarray}
O_{1} &=&({\bar{s}}_{i}c_{j})_{V-A}({\bar{c}}_{j}b_{i})_{V-A}~,  \nonumber \\
O_{2} &=&({\bar{s}}_{i}c_{i})_{V-A}({\bar{c}}_{j}b_{j})_{V-A}~,  \nonumber \\
O_{3} &=&({\bar{s}}_{i}b_{i})_{V-A}\sum_{q}({\bar{q}}_{j}q_{j})_{V-A}~, 
\nonumber \\
O_{4} &=&({\bar{s}}_{i}b_{j})_{V-A}\sum_{q}({\bar{q}}_{j}q_{i})_{V-A}~, 
\nonumber \\
O_{5} &=&({\bar{s}}_{i}b_{i})_{V-A}\sum_{q}({\bar{q}}_{j}q_{j})_{V+A}~, 
\nonumber \\
O_{6} &=&({\bar{s}}_{i}b_{j})_{V-A}\sum_{q}({\bar{q}}_{j}q_{i})_{V+A}~, 
\nonumber \\
O_{7} &=&\frac{em_{b}}{8\pi ^{2}}{\bar{s}}_{i}\sigma ^{\mu \nu }(1+\gamma
_{5})b_{i}F_{\mu \nu }~,  \nonumber \\
O_{8} &=&\frac{g_{s}m_{b}}{8\pi ^{2}}{\bar{s}}_{i}\sigma ^{\mu \nu
}(1+\gamma _{5})T_{ij}^{a}b_{j}G_{\mu \nu }^{a}~.  \label{02a}
\end{eqnarray}
Here $i,j$ are color indices, and we neglect the CKM element $%
V_{ub}V_{us}^{*}$ as well as the $s$-quark mass. The leading contribution to 
$B\rightarrow K_{1}\gamma $ comes from the electromagnetic operator $O_{7}$
as shown in Fig.\ a.As in the case of the real photon emission ($q^{2}=0$),
the only form factor appears in the calculation is $\xi _{\bot }^{(K_{1})}$.
Therefore one can write 
\begin{eqnarray}
\langle O_{7}\rangle _{A} &\equiv &\langle K_{1}(p^{\prime },\epsilon
)\gamma (q,e)|O_{7}|B(p)\rangle  \nonumber \\
&=&\frac{em_{b}}{4\pi ^{2}}\xi _{\bot }^{(K_{1})}\left[ \epsilon ^{*}\cdot
q(p+p^{\prime })\cdot e^{*}-\epsilon ^{*}\cdot e^{*}(p^{2}-p^{\prime
2})+i\epsilon _{\mu \nu \alpha \beta }e^{*\mu }\epsilon ^{*\nu }q^{\alpha
}(p+p^{\prime })^{\beta }\right] ,  \label{03}
\end{eqnarray}
with $\epsilon ^{*\nu }$ and $e^{\mu }$ being the polarization vector for
axial kaon and the photon respectively. The decay rate is straightforwardly
obtained to be\cite{Lee} 
\begin{equation}
\Gamma (B\rightarrow K_{1}\gamma )=\frac{G_{F}^{2}\alpha m_{b}^{2}m_{B}^{3}}{%
32\pi ^{4}}|V_{tb}V_{ts}^{*}|^{2}\Bigg(1-\frac{m^{2}}{m_{B}^{2}}\Bigg)%
^{3}|\xi _{\bot }^{(K_{1})}|^{2}|C_{7}^{{\rm eff(0)}}|^{2}~,  \label{04}
\end{equation}
where $\alpha $ is the fine-structure constant and $C_{7}^{{\rm eff(0)}}$ is
the effective Wilson coefficient at leading order.

\section{Matrix Elements at Next-To-Leading Order of $O(\alpha _{s})$}

At next to leading order of $\alpha _{s}$, there are the contributions from
the operators $O_{2}$ and $O_{8}$ along with that of the $O_{7}$ in $%
B\rightarrow K_{1}\gamma $ decay. Each operator has its vertex contribution
and hard spectator contribution terms which we calculate explicitly.

\subsection{Hard Spectator Contribution}

The Hard spectator contribution is well described by the convolution between
the hard kernel $T_{k}$ and the light cone distribution amplitudes of the
involved mesons, $\Phi _{B}$ and $\Phi _{K_{1}}$ and can be written as $\Phi
_{B}\otimes T_{k}\otimes \Phi _{K_{1}}$. The corresponding decay amplitude
can be calculated in the form of convolution formula, whose leading term can
be expressed as\cite{Ali} 
\begin{eqnarray}
\Delta {\cal M}^{{\rm (HSA)}} &=&\frac{4\pi \alpha _{s}C_{F}}{N_{c}}%
\,\int\limits_{0}^{1}du\int\limits_{0}^{\infty
}dl_{+}\,M_{jk}^{(B)}M_{li}^{(\rho )}{\cal T}_{ijkl}  \nonumber \\
&&  \label{05}
\end{eqnarray}
where $N_{c}$ is the number of colors, $C_{F}=(N_{c}^{2}-1)/(2N_{c})$ is the
Casimir operator eigenvalue in the fundamental representation of the color $%
SU(N_{c})$ group. The leading-twist two-particle light-cone projection
operators $M_{jk}^{(B)}$~\cite{Ball:1998sk,Beneke:2001wa} and $%
M_{li}^{(K_{1})}$~\cite{Grozin:1997pq,Beneke:2001wa} of~$B$- and~$K_{1}$%
-mesons in the momentum representation are: 
\begin{eqnarray}
\hspace*{-10mm} &&M_{jk}^{(B)}=-\frac{if_{B}M}{4}\left. \left[ \frac{1+v%
\hskip-0.45em/}{2}\left\{ \phi _{+}^{(B)}(l_{+})n\hskip-0.45em/_{+}+\phi
_{-}^{(B)}(l_{+})\left( n\hskip-0.45em/_{-}-l_{+}\gamma _{\perp }^{\mu }%
\frac{\partial }{\partial l_{\perp }^{\mu }}\right) \right\} \gamma
_{5}\right] _{jk}\right| _{l=(l_{+}/2)n_{+}}\hspace*{-2mm},  \label{eq:PrO-B}
\\
\hspace*{-10mm} &&M_{li}^{(K_{1})}=-\frac{i}{4}\left[ f_{\perp
}^{(K_{1})}\,\left( \varepsilon \hskip-0.45em/^{*}p\hskip-0.45em/\,\right)
\gamma _{5}\phi _{\perp }^{(K_{1})}(u)+f_{\Vert }^{(K_{1})}\left( \,p\hskip%
-0.45em/\,\frac{m}{E}\,(v\varepsilon ^{*})\,\right) \gamma _{5}\phi _{\Vert
}^{(K_{1})}(u)\right] _{li},  \label{eq:PrO-rho}
\end{eqnarray}
where $f_{B}$ is the $B$-meson decay constant, $f_{\Vert }^{(K_{1})}$ and~$%
f_{\perp }^{(K_{1})}$ are the longitudinal and transverse $K_{1}$-meson
decay constants, respectively, and $\varepsilon _{\mu }$ is the $K_{1}$%
-meson polarization vector. These projectors include also the leading-twist
distribution amplitudes~$\phi _{+}^{(B)}(l_{+})$ and~$\phi _{-}^{(B)}(l_{+})$
of the $B$-meson and~$\phi _{\Vert }^{(K_{1})}(u)$ and~$\phi _{\perp
}^{(K_{1})}(u)$ of the $K_{1}$-meson. ${\cal T}_{ijkl}$ is the
hard-scattering amplitude.The Kinematical relations are used to calculate
the hard spectator contributions are\cite{Beneke:2001wa} 
\begin{eqnarray*}
p_{b}^{\mu } &\simeq &m_{b}\,v^{\mu },\qquad l^{\mu }=\frac{l_{+}}{2}%
\,n_{+}^{\mu }+l_{\perp }^{\mu }+\frac{l_{-}}{2}\,n_{-}^{\mu } \\
k_{1}^{\mu } &\simeq &u\,E\,n_{-}^{\mu }+k_{\perp }^{\mu }+O(k_{\perp
}^{2}),\qquad k_{2}^{\mu }\simeq \bar{u}\,E\,n_{-}^{\mu }-k_{\perp }^{\mu
}+O(k_{\perp }^{2}), \\
v^{2} &=&1,\qquad v^{\mu }=(n_{-}^{\mu }+n_{+}^{\mu })/2\qquad E\simeq M/2 \\
q^{\mu } &=&\omega n_{+}^{\mu }\qquad \omega =M/2
\end{eqnarray*}
To calculate ${\cal T}_{ijkl}$ let's consider the contribution from all the
possible diagrams as done for the $B\rightarrow V\gamma $ \cite{Ali}.

\subsubsection{Spectator corrections due to the electromagnetic dipole
operator $O_{7}$}

The corresponding diagrams are presented in Fig. 1 and the explicit
expression is given by 
\begin{eqnarray}
{\cal T}_{ijkl}^{(1)} &=&-i\frac{G_{F}}{\sqrt{2}}V_{td}^{*}V_{tb}C_{7}^{%
\text{eff}}(\mu )\frac{em_{b}(\mu )}{4\pi ^{2}}\frac{[\gamma _{\mu }]_{kl}}{%
\left( l-k_{2}\right) ^{2}}  \nonumber \\
&&\times \left[ (q\sigma e^{*})(1+\gamma _{5})\frac{\not{p}_{b}+\not{l}-\not%
{k}_{2}+m_{b}}{(p_{b}+l-k_{2})^{2}-m_{b}^{2}}\gamma _{\mu }\right.  \nonumber
\\
&&\left. +\gamma _{\mu }\frac{\not{k}_{1}+\not{k}_{2}-\not{l}}{%
(k_{1}+k_{2}+l)^{2}}(q\sigma e^{*})(1+\gamma _{5})\right] _{ij}  \label{3.1}
\end{eqnarray}
where the short hand notation is used for $(q\sigma e^{*})=\sigma ^{\mu \nu
}q_{\mu }e_{\nu }^{*}$.

\subsubsection{Spectator corrections due to the chromomagnetic dipole
operator $O_{8}$}

The corresponding diagrams are presented in Fig. 2. The first two diagrams
(Fig. 2a) show the corrections for the case when the photon is emitted from
the flavor changing quark line and the result is 
\begin{eqnarray}
{\cal T}_{ijkl}^{(2a)} &=&-i\frac{G_{F}}{\sqrt{2}}V_{td}^{*}V_{tb}C_{8}^{%
\text{eff}}(\mu )\frac{em_{b}(\mu )}{4\pi ^{2}}[\gamma _{\nu }]_{kl}\frac{%
\left( l-k_{2}\right) _{\mu }}{\left( l-k_{2}\right) ^{2}}  \nonumber \\
&&\times \left[ \not{e}^{*}\frac{\not{p}_{b}+\not{l}-\not{k}_{2}}{%
(p_{b}+l-k_{2})^{2}}\sigma _{\mu \nu }(1+\gamma _{5})\right.  \nonumber \\
&&+\left. \sigma _{\mu \nu }(1+\gamma _{5})\frac{\not{k}_{1}+\not{k}_{2}-\not%
{l}+m_{b}}{(k_{1}+k_{2}+l)^{2}-m_{b}^{2}}\not{e}^{*}\right] _{ij}
\label{3.2}
\end{eqnarray}
Fig. 2b contains the diagrams with the photon emission from the spectator
quark which results into the following hard-scattering amplitude: 
\begin{eqnarray}
{\cal T}_{ijkl}^{(2b)} &=&i\frac{G_{F}}{\sqrt{2}}V_{td}^{*}V_{tb}C_{8}^{%
\text{eff}}(\mu )\frac{eQ_{d[u]}m_{b}(\mu )}{4\pi ^{2}}  \nonumber \\
&&\times \left[ \sigma _{\mu \nu }(1+\gamma _{5})\right] _{ij}\frac{\left(
p_{b}-k_{1}\right) _{\mu }}{\left( p_{b}-k_{1}\right) ^{2}}  \nonumber \\
&&\times \left[ \gamma _{\nu }\frac{\not{p}_{b}+\not{l}-\not{k}_{1}}{%
(p_{b}+l-k_{1})^{2}}\not{e}^{*}+\not{e}^{*}\frac{\not{k}_{1}+\not{k}_{2}-\not%
{p}_{b}}{(k_{1}+k_{2}-p_{b})^{2}}\gamma _{\nu }\right] _{kl}  \label{3.3}
\end{eqnarray}
where $Q_{d[u]}$ is the charge of the spectator quark.

\subsubsection{Spectator corrections involving the penguin-type diagrams and
the operator $O_{2}$}

The corresponding diagrams are shown in Figs. 3, 4 and 5. The hard spectator
contribution corresponding to the diagrams in Fig. 3a is 
\begin{eqnarray}
{\cal T}_{ijkl}^{(3a)} &=&\frac{G_{F}}{\sqrt{2}}\frac{e}{24\pi ^{2}}%
\sum_{f=u,c}V_{fd}^{*}V_{fb}C_{2}^{(f)}(\mu )\Delta F_{1}\left(
z_{1}^{(f)}\right) [\gamma _{\nu }]_{kl}  \nonumber \\
&&\times \left[ \left\{ \gamma _{\nu }-\frac{\left( k_{2}-l\right) _{\nu
}\left( \not{k}_{2}-\not{l}\right) }{\left( k_{2}-l\right) ^{2}}\right\}
(1-\gamma _{5})\frac{\not{k}_{1}+\not{k}_{2}-\not{l}+m_{b}}{%
(k_{1}+k_{2}+l)^{2}-m_{b}^{2}}\not{e}^{*}\right.  \nonumber \\
&&+\left. \not{e}^{*}\frac{\not{p}_{b}+\not{l}-\not{k}_{2}}{%
(p_{b}+l-k_{2})^{2}}\left\{ \gamma _{\nu }-\frac{\left( k_{2}-l\right) _{\nu
}\left( \not{k}_{2}-\not{l}\right) }{\left( k_{2}-l\right) ^{2}}\right\}
(1-\gamma _{5})\right] _{ij}  \label{3.4}
\end{eqnarray}
and from the diagrams in Fig. 3b, where the photon is emitted from the
spectator quark line yield: 
\begin{eqnarray}
{\cal T}_{ijkl}^{(3b)} &=&\frac{G_{F}}{\sqrt{2}}\frac{eQ_{d[u]}}{24\pi ^{2}}%
\sum_{f=u,c}V_{fd}^{*}V_{fb}C_{2}^{(f)}(\mu )\Delta F_{1}\left(
z_{0}^{(f)}\right)  \nonumber \\
&&\times \left[ \not{e}^{*}\frac{\not{k}_{1}+\not{k}_{2}-\not{p}_{b}}{%
(k_{1}+k_{2}-p_{b})^{2}}\gamma _{\nu }+\gamma _{\nu }\frac{\not{p}_{b}+\not%
{l}-\not{k}_{1}}{(p_{b}+l-k_{1})^{2}}\not{e}^{*}\right] _{kl}  \nonumber \\
&&\times \left[ \left\{ \gamma _{\nu }-\frac{\left( p_{b}-k_{1}\right) _{\nu
}}{\left( p_{b}-k_{1}\right) ^{2}}\left( \not{p}_{b}-\not{k}_{1}\right)
\right\} (1-\gamma _{5})\right] _{ij}  \label{3.5}
\end{eqnarray}
The detailed discussion about $\Delta F_{1}\left( z_{1}^{(f)}\right) $ and $%
\Delta F_{1}\left( z_{0}^{(f)}\right) $ can be found in\cite{Ali}.

The contributions from the diagrams in Fig. 4 can be written as 
\begin{eqnarray}
{\cal T}_{ijkl}^{(4)} &=&-\frac{G_{F}}{\sqrt{2}}\frac{e}{6\pi ^{2}}\frac{%
[\gamma _{\nu }]_{kl}}{\left( k_{2}-l\right) ^{2}\left( q\left[
k_{2}-l\right] \right) }\sum_{f=u,c}V_{fd}^{*}V_{fb}C_{2}^{(f)}(\mu ) 
\nonumber \\
&&\times \left[ \left\{ 
\begin{array}{c}
\left[ 
\begin{array}{c}
q_{\nu }\text{E}\left( k_{2}-l,e^{*},q\right) -\left( q\left[ k_{2}-l\right]
\right) \text{E}\left( \nu ,e^{*},q\right) \\ 
+\left( e^{*}\left[ k_{2}-l\right] \right) \text{E}\left( q,\nu
,k_{2}-l\right) \\ 
-\left( q\left[ k_{2}-l\right] \right) \text{E}\left( e^{*},\nu
,k_{2}-l\right)
\end{array}
\right] \Delta i_{5}\left( z_{0}^{(f)},z_{1}^{(f)},0\right) \\ 
+\left[ 
\begin{array}{c}
\left( k_{2}-l\right) ^{2}\text{E}\left( \nu ,e^{*},q\right) \\ 
+\left( k_{2}-l\right) _{\nu }\text{E}\left( e^{*},k_{2}-l,q\right)
\end{array}
\right] \Delta i_{25}\left( z_{0}^{(f)},z_{1}^{(f)},0\right)
\end{array}
\right\} \left( 1-\gamma _{5}\right) \right] _{ij}  \label{3.6}
\end{eqnarray}
where 
\begin{equation}
{\rm E}(\mu ,\nu ,\rho )\equiv \frac{1}{2}\,(\gamma _{\mu }\gamma _{\nu
}\gamma _{\rho }-\gamma _{\rho }\gamma _{\nu }\gamma _{\mu })=-i\varepsilon
_{\mu \nu \rho \sigma }\,\gamma ^{\sigma }\,\gamma _{5}.  \label{3.7}
\end{equation}
and the form of $\Delta i_{5}\left( z_{0}^{(f)},z_{1}^{(f)},0\right) $ and $%
\Delta i_{25}\left( z_{0}^{(f)},z_{1}^{(f)},0\right) $ along with the
detailed discussion is given in\cite{Ali}.

Finally, there are the diagrams where the photon is emitted from the
internal quark line due to the effective $b\rightarrow s\gamma $ interaction
and a gluon is exchanged between the spectator quark and the $b$- or $s$
quark as shown in Fig. 5. For on shell photon such kind of diagrams do not
contribute and hence the contribution comes from the Fig. 5. is zero.

Using Equations (\ref{eq:PrO-B}) and (\ref{eq:PrO-rho}) along with the hard
scattering matrix derived in the Eqs. (\ref{3.1}-\ref{3.6}), we can write
from Eq. (\ref{05}) as 
\begin{eqnarray}
\Delta {\cal M}_{{\rm sp}}^{(K_{1})} &=&\frac{G_{F}}{\sqrt{2}}\,\frac{%
e\alpha _{s}C_{F}}{4\pi N_{c}}\,f_{B}f_{\perp }^{(K_{1})}M\,\left[
(e^{*}\varepsilon ^{*})+i\,{\rm eps}(e^{*},\varepsilon ^{*},n_{-},v)\right]
\,\sum_{k=1}^{5}\Delta H_{k}^{(K_{1})}  \nonumber \\
&&  \label{main}
\end{eqnarray}
where ${\rm eps}(a,b,c,d)=\varepsilon _{\mu \nu \rho \sigma }a^{\mu }b^{\nu
}c^{\rho }d^{\sigma }$ and the upper index~$K_{1}$ characterizes the final
axial meson. The dimensionless functions~$\Delta H_{k}^{(K_{1})}$ ($%
k=1,2,3,4,5$) describe the contributions of the sets of Feynman diagrams
presented in Figs.~1-5, respectively. In the leading order of the inverse $B$%
-meson mass ($\sim \Lambda _{{\rm QCD}}/M$), the result reads as follows: 
\begin{eqnarray}
\Delta H_{1}^{(K_{1})}(\mu ) &\simeq &V_{ts}^{*}V_{tb}\,C_{7}^{{\rm eff}%
}(\mu )\,m_{b}(\mu )\left[ \left\langle l_{+}^{-1}\right\rangle
_{+}\left\langle \bar{u}^{-1}\right\rangle _{\perp }^{(K_{1})}(\mu
)+\left\langle l_{+}^{-1}\right\rangle _{-}\left\langle \bar{u}%
^{-2}\right\rangle _{\perp }^{(K_{1})}(\mu )\right] ,  \label{3.8a} \\
\Delta H_{2}^{(K_{1})}(\mu ) &\simeq &\frac{1}{3}\,V_{ts}^{*}V_{tb}\,C_{8}^{%
{\rm eff}}(\mu )\,m_{b}(\mu )\,\left\langle l_{+}^{-1}\right\rangle
_{+}\left\langle u^{-1}\right\rangle _{\perp }^{(K_{1})}(\mu ),  \label{3.9}
\\
\Delta H_{3}^{(K_{1})}(\mu ) &\simeq &0,  \label{3.10} \\
\Delta H_{4}^{(K_{1})}(\mu ) &\simeq &\frac{1}{3}\,C_{2}(\mu
)\,M\,\left\langle l_{+}^{-1}\right\rangle _{+}\left[
V_{ts}^{*}V_{tb}\,\left\langle \bar{u}^{-1}\right\rangle _{\perp
}^{(K_{1})}(\mu )+V_{cs}^{*}V_{cb}\,h^{(K_{1})}(z,\mu )\right] ,
\label{3.11} \\
\Delta H_{5}^{(K_{1})}(\mu ) &\simeq &0,  \label{3.12}
\end{eqnarray}
where $z=m_{c}^{2}/m_{b}^{2}$ and the short-hand notation used are for the
integrals over the mesons distribution functions: 
\begin{equation}
\left\langle l_{+}^{N}\right\rangle _{\pm }\equiv \int\limits_{0}^{\infty
}dl_{+}\,l_{+}^{N}\,\phi _{\pm }^{(B)}(l_{+}),\qquad \left\langle
f\right\rangle _{\perp ,\Vert }^{(K_{1})}(\mu )\equiv
\int\limits_{0}^{1}du\,f(u)\,\phi _{\perp ,\Vert }^{(K_{1})}(u,\mu ),
\label{3.13}
\end{equation}
and for convenience the following function is introduced: 
\begin{equation}
h^{(K_{1})}(z,\mu )=\left\langle \frac{\Delta i_{5}(z_{0}^{(c)},0,0)+1}{\bar{%
u}}\right\rangle _{\!\!\perp }^{\!\!(K_{1})}.  \label{3.14}
\end{equation}
The expressions of $\Delta H_{k}^{(K_{1})}$ given in Eqs. (\ref{3.8a}-\ref
{3.12}) are similar to those obtained for $B\rightarrow K^{*}\gamma $ (c.f.
Eqs. (4.4-4.6) of Ali et al.\cite{Ali}) which show that the additional $%
\gamma _{5}\,$present in the DA of $K_{1}$ has no effect on the
calculations. Using the above Equations in Eq. (\ref{main}) one can write 
\begin{eqnarray}
\Delta {\cal M}_{{\rm sp}} &=&\frac{G_{F}}{\sqrt{2}}\,V_{tp}^{*}V_{tb}\,%
\frac{\alpha _{s}C_{F}}{4\pi }\,\frac{e}{4\pi ^{2}}\,\Delta F_{\perp
}^{(K_{1})}(\mu )\left[ (pP)\,(e^{*}\varepsilon ^{*})+i\,{\rm eps}%
(e^{*},\varepsilon ^{*},p,P)\right]  \nonumber \\
&\times &\left[ C_{7}^{{\rm eff}}(\mu )+\frac{1}{3}\,C_{8}^{{\rm eff}}(\mu
)\,\frac{\left\langle u^{-1}\right\rangle _{\perp }^{(K_{1})}}{\left\langle 
\bar{u}^{-1}\right\rangle _{\perp }^{(K_{1})}}+\frac{1}{3}\,C_{2}(\mu
)\left( 1+\frac{V_{cs}^{*}V_{cb}}{V_{ts}^{*}V_{tb}}\,\frac{h^{(K_{1})}(z,\mu
)}{\left\langle \bar{u}^{-1}\right\rangle _{\perp }^{(K_{1})}(\mu )}\right)
\right] ~,  \label{3.15}
\end{eqnarray}
where 
\begin{equation}
\Delta F_{\perp }^{(K_{1})}(\mu )=\frac{8\pi ^{2}f_{B}f_{\perp
}^{(K_{1})}(\mu )}{N_{c}M\lambda _{B,+}}\left\langle \bar{u}%
^{-1}\right\rangle _{\perp }^{(K_{1})}(\mu ),  \label{3.16}
\end{equation}
is the dimensionless quantity. $\lambda _{B,+}^{-1}=\left\langle
l_{+}^{-1}\right\rangle _{+}$ is the first negative moment of the $B$-meson
distribution function~$\phi _{+}^{(B)}(l_{+})$ which is typically estimated
as $\lambda _{B,+}^{-1}=(3\pm 1)$~GeV~\cite{Grozin:1997pq,Beneke:2001wa}. In
a recent paper by Braun et al. \cite{Braun}, the scale dependence of this
moment is worked out at next to leading order and the value obtained is $%
\lambda _{B,+}^{-1}(1$ GeV$)=(2.15\pm 0.50)$~GeV. At the scale~$\mu _{{\rm sp%
}}=\sqrt{\mu _{b}\Lambda _{H}}$ of the hard-spectator corrections, and for
the central values of the parameters are shown in Table I.

The analytical expression for the function $h^{(V)}(z,\mu )$ for the vector
meson is given in\cite{Ali}. We will proceed to give an analytical result
for the axial meson. One can write the leading twist distribution amplitude $%
\phi _{\perp }^{(K_{1})}(u,\mu )$ as\cite{Ball:1998sk} 
\begin{equation}
\phi _{\perp }^{(K_{1})}(u,\mu )=6u\bar{u}\left[ 1+\sum_{n=1}^{\infty
}a_{\perp n}^{(K_{1})}(\mu )\,C_{n}^{3/2}(u-\bar{u})\right] ,  \label{3.17}
\end{equation}
where $C_{n}^{3/2}(u-\bar{u})$ are the Gegenbauer polynomials [$%
C_{1}^{3/2}(u-\bar{u})=3(u-\bar{u})$, $C_{2}^{3/2}(u-\bar{u})=3\left[ 5(u-%
\bar{u})^{2}-1\right] $/2, etc.] and $a_{\perp n}^{(K_{1})}(\mu )$ are the
corresponding Gegenbauer moments. These moments are scale dependent and so
should be evaluated at the scale~$\mu $; their scale dependence is governed
by\cite{Ball:1998sk}: 
\begin{equation}
a_{\perp n}^{(K_{1})}(\mu )=\left( \frac{\alpha _{s}(\mu ^{2})}{\alpha
_{s}(\mu _{0}^{2})}\right) ^{\gamma _{n}/\beta _{0}}a_{\perp
n}^{(K_{1})}(\mu _{0}),\qquad \gamma _{n}=4C_{F}\left( \sum_{k=1}^{n}\frac{1%
}{k}-\frac{n}{n+1}\right) ,  \label{3.18}
\end{equation}
where $\beta _{0}=(11N_{c}-2n_{f})/3$ and~$\gamma _{n}$ is the one-loop
anomalous dimension with $C_{F}=(N_{c}^{2}-1)/(2N_{c})=4/3$. In the limit~$%
\mu \to \infty $ the Gegenbauer moments vanish,~$a_{\perp n}^{(K_{1})}(\mu
)\to 0$, and the leading-twist transverse distribution amplitude has its
asymptotic form: 
\begin{equation}
\phi _{\perp }^{(K_{1})}(u,\mu )\to \phi _{\perp }^{({\rm as})}(u)=6u\bar{u}.
\label{3.19}
\end{equation}
A simple model of the transverse distribution which includes contributions
from the first~$a_{\perp 1}^{(K_{1})}(\mu )$ and the second~$a_{\perp
2}^{(K_{1})}(\mu )$ Gegenbauer moments only is used here in the analysis. In
this approach the quantities~$<u^{-1}>_{\perp }^{(K_{1})}$ and~$<\bar{u}%
^{-1}>_{\perp }^{(K_{1})}$ are: 
\begin{equation}
<u^{-1}>_{\perp }^{(K_{1})}=3\left[ 1-a_{\perp 1}^{(K_{1})}(\mu )+a_{\perp
2}^{(K_{1})}(\mu )\right] ,\qquad <\bar{u}^{-1}>_{\perp }^{(K_{1})}=3\left[
1+a_{\perp 1}^{(K_{1})}(\mu )+a_{\perp 2}^{(K_{1})}(\mu )\right] ,
\label{3.20}
\end{equation}
and depend on the scale~$\mu $ due to the coefficients~$a_{\perp
n}^{(K_{1})}(\mu )$. The calculation for the axial $K$ meson without
Gegenbauer moments is done in detail\cite{Lee,Leenew}. In our calculation we
will incorporate these effects in the calculations and will check the
sensitivity of branching ratio with the LEET form factors in the presence of
these moments. The Gegenbauer moments were evaluated at the scale $\mu
_{0}=1 $~GeV, yielding~\cite{Ball:1998sk}: $a_{\perp 1}^{(K^{*})}(1~{\rm GeV}%
)=0.20\pm 0.05$ and $a_{\perp 2}^{(K^{*})}(1~{\rm GeV})=0.04\pm 0.04$ for
the $K^{*}$-meson. The value of these two Gengenbauer moments have recently
been modified and it has been pointed out that these values are now larger
in magnitude, have larger errors and, moreover, the first Gegenbauer moment
changes it sign\cite{Ball}. The new values of these Gengenbauer moments are $%
a_{\perp 1}^{(K^{*})}(1~{\rm GeV})=-0.34\pm 0.18$ and $a_{\perp
2}^{(K^{*})}(1~{\rm GeV})=0.13\pm 0.08$ for the $K^{*}$-meson\cite
{Ali-Lunghi}. We will use the same value for the $K_{1}$ because one can see
that the value is not changed for the axial meson also because changing the
scale has not the noticeable effect on the coupling constants and so on the
Gengenbauer moments\cite{Ball:1998sk}, \cite{Patricia}. In the same manner,
the function~$h^{(K_{1})}(z,\mu )$ introduced in Eq.~(\ref{3.15}) can be
presented as an expansion on the Gegenbauer moments: 
\begin{eqnarray}
h^{(K_{1})}(z,\mu ) &=&h_{0}(z)+a_{\perp 1}^{(K_{1})}(\mu
)\,h_{1}(z)+a_{\perp 2}^{(K_{1})}(\mu )\,h_{2}(z)  \label{3.21} \\
&=&\left[ 1+3a_{\perp 1}^{(K_{1})}(\mu )+6a_{\perp 2}^{(K_{1})}(\mu )\right]
\left\langle (\Delta i_{5}+1)/\bar{u}\right\rangle _{\perp }^{(0)}  \nonumber
\\
&-&6\left[ a_{\perp 1}^{(K_{1})}(\mu )+5a_{\perp 2}^{(K_{1})}(\mu )\right]
\left\langle \Delta i_{5}+1\right\rangle _{\perp }^{(0)}+30\,a_{\perp
2}^{(K_{1})}(\mu )\left\langle \bar{u}\,(\Delta i_{5}+1)\right\rangle
_{\perp }^{(0)},  \nonumber
\end{eqnarray}
where another short-hand notation is introduced for the integral: 
\begin{equation}
\left\langle f(u)\right\rangle _{\perp
}^{(0)}=\int\limits_{0}^{1}du\,f(u)\,\phi _{\perp }^{({\rm as})}(u).
\label{3.22}
\end{equation}
The detail of relevant functions as well as the analytical form of the $%
\left\langle (\Delta i_{5}+1)/\bar{u}\right\rangle _{\perp }^{(0)}$, $%
\left\langle \Delta i_{5}+1\right\rangle _{\perp }^{(0)}$ and $\left\langle 
\bar{u}\,(\Delta i_{5}+1)\right\rangle _{\perp }^{(0)}$ is given\cite{Ali}.
The real and imaginary parts of the functions~$h_{n}(z)$ are presented in
Figs.~6 (for $n=0$) and~7 (for $n=1$ and $n=2$). The dependence on $%
z=m_{c}^{2}/m_{b}^{2}$ of the function $h^{(K_{1})}(z,\mu )$ at the mass
scale $\mu =\mu _{\text{sp}}=1.52$ GeV of hard-spectator corrections is
presented in Fig.8. We have observed that our plots given in Fig. 7 and Fig.
8 are different to those given by Ali et al. (c.f. Fig. 7 and Fig. 8)\cite
{Ali}. The authors of the article \cite{Ali} are agree to this observation 
\cite{Aliprivate}, pointed out by Gilani \cite{0404026}. The value of the
corresponding Gegenbauer moments used for the evaluation are given in Table
I.

\begin{center}
\begin{tabular}{l}
\ \ \ \ \ \ \ \ \ \ \ \ \ \ \ \ \ \ \ \ \ \ \ \ \ \ \ \ \ \ \ \ \ \ \ \ \ $%
\,\,\,\,\,\,\,K_{1}(1270)$\ \ \ \ \ \ \ \ \ \ \ \thinspace \thinspace
\thinspace \thinspace \thinspace \thinspace \thinspace \thinspace \thinspace
\thinspace \thinspace \thinspace \thinspace \thinspace \thinspace \thinspace
\thinspace \thinspace \thinspace \thinspace \thinspace \thinspace \thinspace
\thinspace \thinspace \thinspace \thinspace \thinspace \thinspace \thinspace
\thinspace \thinspace \thinspace \thinspace \thinspace $\,\,K_{1}(1400)$\ \
\ \ \ \ \ \ \ \ \ \ \ \ \ \ \ \ \ \ \ \ \ \thinspace
\end{tabular}

\thinspace \thinspace \thinspace \thinspace \thinspace \thinspace \thinspace
\thinspace \thinspace \thinspace \thinspace \thinspace \thinspace \thinspace
\thinspace \thinspace 
\begin{tabular}{lllll}
& $\,\,\,\,\mu _{\text{sp}}$ & $\,\,\,m_{b\text{,pole}}$ & $\,\,\,\mu _{%
\text{sp}}$ & $\,\,\,\,m_{b\text{,pole}}$ \\ 
$\mu $, [GeV]\thinspace \thinspace \thinspace & $\,\,\,1.52$ & $\,\,\,4.65$
& $\,\,\,1.52$ & $\,\,\,4.65$ \\ 
$a_{\bot _{1}}(\mu )$ & $-0.321$ & $-0.285$ & $-0.321$ & $-0.285$ \\ 
$a_{\bot _{2}}(\mu )$ & $\,\,\,0.118$ & $\,\,\,0.097$ & $\,\,\,0.118$ & $%
\,\,\,0.097$ \\ 
$h_{0}(z)$ & $\,\,\,3.91+i1.64$ & $\,\,\,3.91+i1.64$ & $\,\,\,3.91+i1.64$ & $%
\,\,\,3.91+i1.64$ \\ 
$h^{(V)}(z,\mu )$ & $\,\,\,2.89+i1.74$ & $\,\,\,3.07+i1.34$ & $%
\,\,\,2.89+i1.74$ & $\,\,\,3.07+i1.34$ \\ 
$\left\langle \bar{u}^{-1}\right\rangle _{\bot }^{(V)}(\mu )$ & $\,\,\,2.39$
& $\,\,\,2.43$ & $\,\,2.39$ & $\,\,\,2.43$ \\ 
$h^{(V)}/\left\langle \bar{u}^{-1}\right\rangle _{\bot }^{(V)}$ & $%
\,\,\,1.21+i0.73$ & $\,\,\,1.26+i0.71$ & $1.21+i0.73$ & $\,\,\,1.26+i0.71$
\\ 
$f_{\bot }^{(V)}(\mu )$, [MeV] & $\,\,\,118.6$ & $\,\,\,111.78$ & $%
\,\,\,88.44$ & $\,\,\,83.38$ \\ 
$\Delta F_{\bot }^{(V)}(\mu )$ & $\,\,\,0.55$ & $\,\,\,0.52$ & $\,\,\,0.41$
& $\,\,\,0.39$%
\end{tabular}
\end{center}

The amplitude (\ref{main}) is proportional to the tensor decay constant~$%
f_{\perp }^{(K_{1})}$ of the axial meson which is a scale dependent
parameter. As for the Gegenbauer moments~$a_{\perp n}^{(K_{1})}$, there
values were defined at the mass scale $\mu _{0}=1$~GeV from the LCSR is\cite
{Safir}: $f_{\perp }^{(K_{1})}(1~{\rm GeV})=122$~MeV. Their values at an
arbitrary scale~$\mu $ can be obtained with the help of the evolution
equation~\cite{Ball:1998sk}: 
\begin{equation}
f_{\perp }^{(V)}(\mu )=\left( \frac{\alpha _{s}(\mu ^{2})}{\alpha _{s}(\mu
_{0}^{2})}\right) ^{4/(3\beta _{0})}f_{\perp }^{(V)}(\mu _{0}).  \label{3.23}
\end{equation}
Central values of the tensor decay constants at the scales $\mu _{{\rm sp}%
}=1.52$~GeV and $m_{b,{\rm pole}}=4.65$~GeV are presented in Table I.

\subsubsection{Branching Ratio for $B\rightarrow K_{1}\gamma $}

The branching ratio for $B\rightarrow K_{1}\gamma $ is simply given by 
\begin{eqnarray}
{\cal B}_{{\rm th}}(B\to K^{*}\gamma ) &=&\tau _{B}\,\Gamma _{{\rm th}}(B\to
K^{*}\gamma )  \nonumber \\
&=&\tau _{B}\,\frac{G_{F}^{2}\alpha |V_{tb}V_{ts}^{*}|^{2}}{32\pi ^{4}}%
\,m_{b,{\rm pole}}^{2}\,M^{3}\,\left[ \xi _{\perp }^{(K_{1})}\right]
^{2}\left( 1-\frac{m_{K^{*}}^{2}}{M^{2}}\right) ^{3}\left| C_{7}^{(0){\rm eff%
}}+A^{(1)}(\mu )\right| ^{2}  \nonumber \\
&&  \label{branching1}
\end{eqnarray}
where~$G_{F}$ is the Fermi coupling constant, $\alpha =\alpha (0)=1/137$ is
the fine-structure constant, $m_{b,{\rm pole}}$ is the pole $b$-quark mass, $%
M$~and $m_{K_{1}}$ are the $B$- and $K_{1}$-meson masses, and~$\tau _{B}$ is
the lifetime of the~$B^{0}$- or $B^{+}$-meson. The value of these constants
is used from\cite{Ali} for the numerical analysis. For this study, we
consider $\xi _{\perp }^{(K_{1})}$ as a free parameter and we will extract
its value from the current experimental data on $B\to K_{1}\gamma $ decays.

The function~$A^{(1)}$ in Eq.~(\ref{branching1}) can be decomposed into the
following three components: 
\begin{equation}
A^{(1)}(\mu )=A_{C_{7}}^{(1)}(\mu )+A_{{\rm ver}}^{(1)}(\mu )+A_{{\rm sp}%
}^{(1)K_{1}}(\mu _{{\rm sp}})~.  \label{eq:A1tb}
\end{equation}
Here, $A_{C_{7}}^{(1)}$ and $A_{{\rm ver}}^{(1)}$ are the $O(\alpha _{s})$
(i.e. NLO) corrections due to the Wilson coefficient~$C_{7}^{{\rm eff}}$ and
in the $b\to s\gamma $ vertex, respectively, and $A_{{\rm sp}}^{(1)K_{1}}$
is the ${\cal O}(\alpha _{s})$ hard-spectator corrections to the $B\to
K_{1}\gamma $ amplitude computed in this paper. Their explicit expressions
are as follows: 
\begin{eqnarray}
A_{C_{7}}^{(1)}(\mu ) &=&\frac{\alpha _{s}(\mu )}{4\pi }\,C_{7}^{(1){\rm eff}%
}(\mu ),  \label{eq:A1tb-C7} \\
A_{{\rm ver}}^{(1)}(\mu ) &=&\frac{\alpha _{s}(\mu )}{4\pi }\left\{ \frac{32%
}{81}\left[ 13C_{2}^{(0)}(\mu )+27C_{7}^{(0){\rm eff}}(\mu )-9\,C_{8}^{(0)%
{\rm eff}}(\mu )\right] \ln \frac{m_{b}}{\mu }\right.  \label{eq:A1tb-ver} \\
&-&\left. \frac{20}{3}\,C_{7}^{(0){\rm eff}}(\mu )+\frac{4}{27}\left(
33-2\pi ^{2}+6\pi i\right) C_{8}^{(0){\rm eff}}(\mu
)+r_{2}(z)\,C_{2}^{(0)}(\mu )\right\} ,\qquad  \nonumber \\
A_{{\rm sp}}^{(1)K_{1}}(\mu _{{\rm sp}}) &=&\frac{\alpha _{s}(\mu _{{\rm sp}%
})}{4\pi }\,\frac{2\Delta F_{\perp }^{(K_{1})}(\mu _{{\rm sp}})}{9\xi
_{\perp }^{(K_{1})}}\left\{ 3C_{7}^{(0){\rm eff}}(\mu _{{\rm sp}})\right.
\label{eq:A1tb-sp} \\
&+&\left. C_{8}^{(0){\rm eff}}(\mu _{{\rm sp}})\left[ 1-\frac{6a_{\perp
1}^{(K_{1})}(\mu _{{\rm sp}})}{\left\langle \bar{u}^{-1}\right\rangle
_{\perp }^{(K_{1})}(\mu _{{\rm sp}})}\right] +C_{2}^{(0)}(\mu _{{\rm sp}%
})\left[ 1-\frac{h^{(K_{1})}(z,\mu _{{\rm sp}})}{\left\langle \bar{u}%
^{-1}\right\rangle _{\perp }^{(K_{1})}(\mu _{{\rm sp}})}\right] \right\} . 
\nonumber
\end{eqnarray}
Actually $C_{7}^{(1){\rm eff}}(\mu )$ and $A_{{\rm ver}}^{(1)}(\mu )$ are
process independent and encodes the QCD\ effects only, where as $A_{{\rm sp}%
}^{(1)}(\mu _{{\rm sp}})$ contains the key information about the out going
mesons. The factor $\frac{6a_{\perp 1}^{(K_{1})}(\mu _{{\rm sp}})}{%
\left\langle \bar{u}^{-1}\right\rangle _{\perp }^{(K_{1})}(\mu _{{\rm sp}})}$
appearing in the Eq. (\ref{eq:A1tb-sp}) is arising due to the Gegenbauer
moments. The purpose of this paper is to see the effect of these Gegenbauer
moments on the value of the form factor. As it is mentioned in\cite
{Ali,Lee,Leenew}, that the non-asymptotic corrections in the $K_{1}$ meson
wave-function reduces the coefficient of the anomalous choromomagnetic
moment $C_{8}^{(0)\text{eff}}\left( \mu _{\text{sp}}\right) $ by the amount
20\%. Therefore it is viable to calculate the effect of these Gengenbauer
moments. The value obtained for the quantity $\left| C_{7}^{(0){\rm eff}%
}+A^{(1)}(\mu )\right| ^{2}$ at different scales is listed in the TableII
for $K_{1}(1270)$.

\begin{center}
\begin{tabular}{|l|l|l|l|}
\hline
$\,\,\,\,\,\,\,\,\,\,\,\,\,\,\,\,\,\,m_{c}/m_{b}$ & $\,\,\,\,\,\,\,\,\,\,\,%
\,\,\,0.29$ & $\,\,\,\,\,\,\,\,\,\,\,\,\,\,\,\,0.29$ & $\,\,\,\,\,\,\,\,\,\,%
\,\,\,\,\,\,0.29$ \\ \hline
$\,\,\,\,\,\,\,\,\,\,\,\,\,\,\,\,\,\,\,\,\,\,\,\,\,\mu $ & $\bar{m}_{b}=4.27$%
GeV & $m_{b\text{,pole}}=4.65$GeV & $m_{b\text{,PS}}=4.6$GeV \\ \hline
$\left( C_{7}^{(0)\text{eff}}+A^{(1)}(\mu )\right) _{\text{Our}}$ & $%
-0.358-i0.022$ & $-0.356-i0.021$ & $-0.356-i0.021$ \\ \hline
$\left( C_{7}^{(0)\text{eff}}+A^{(1)}(\mu )\right) _{\text{Lee}}$\cite{Lee}
& $-0.406-i0.033$ & $\,\,\,\,\,\,\,\,\,\,\,\,\,\,\,\,\,\,\,\,\,\,\,\,\times $
& $-0.410-i0.033$ \\ \hline
$\left| C_{7}^{(0)\text{eff}}+A^{(1)}(\mu )\right| _{\text{Our}}^{2}$ & $%
0.128$ & $0.127$ & $0.127$ \\ \hline
\end{tabular}

Table II
\end{center}

The numbers given for the quantity $\left( C_{7}^{(0)\text{eff}}+A^{(1)}(\mu
)\right) $ needs some comments. The first one is that with the same value of
the quark mass ratio $\,m_{c}/m_{b}$, the total amplitude has negligible
dependence on the choice of $b$-quark mass or in other words the scale $\mu $%
. Secondly, if the effects of the Gegenbauer moments are included then on
can easily see from third and forth row of the Table II that the value of
the total amplitude reduces as compared to the value given in the literature%
\cite{Lee}.In order to calculate the numerical value for the branching ratio
we use the reference scale to be 
\[
\left( \,\mu \text{, }\mu _{\text{sp}}\right) =\left( 4.27\text{ GeV},\,1.45%
\text{ GeV}\right) 
\]
After calculating the Gengenbauer moments at these scales the only
independent parameter which is left in the calculation of the branching
ratio is the LEET form factor and it has the biggest theoretical
uncertainty.. By taking the value of the form factor from the LCSR which is $%
\xi _{\perp }^{(K_{1})}(0)=0.14\pm 0.03$ it was shown some time ago that the
value of the branching ratio is very small as compared to the experimental
results\cite{Lee}. Then the value of the form factor is extracted from the
experimental measurements (\ref{data2}) and it is found that the value is%
\cite{Leenew} 
\[
\xi _{\perp }^{(K_{1})}(0)=0.32\pm 0.03 
\]
which is much larger than LCSR result for the form factor. and is contrary
to that of the $K^{*}$ meson where the value of LEET form factor $\xi
_{\perp }^{(K^{*})}$ is smaller as compared to the LCSR result. Such
discrepancy in case of $K^{*}$ is not yet known but for the axial $K$ meson
some sources of discrepancies are discussed in\cite{Leenew} and it is said
that it will be bad if the Gengenbauer moments increase the value of this
form factor. But we have shown that it is not the case.

Now by putting the value of the total amplitude at the scale $\left( \,\mu 
\text{, }\mu _{\text{sp}}\right) =\left( 4.27\text{ GeV},\,1.45\text{ GeV}%
\right) $ calculated in the last row of Table II and all the other inputs
from\cite{Lee} in Eq. (\ref{branching1}) one can see that the value
extracted for the form factor remains the same. Thus even if we consider the
non asymptotic form of the light-cone\ DA, it has a very small effect on the
total decay amplitude and leaves all the other things almost same. The
reason is that the dominant contribution comes from the operator $O_{7}$ and
so from the Wilson coefficient $C_{7}$.

In conclusion, we surveyed the implications of the first observation of $%
\,B\rightarrow K_{1}(1270)\gamma $ by using the higher twists in the
light-cone DA that are encoded in the coefficient of Gegenbauer expansion.
By incorporating all these effects the value of the relevant form factor
(LEET) is extracted from the data at NLO of $\alpha _{s}$. It is shown that
the value of the form factor remains the same as calculated with out these
higher twists (asymptotic form). So the non-asymptotic form of DA is not the
suitable candidate to explain the discrepancy..

{\bf Acknowledgments}

One of the authors (J) would like to thank Prof. Fayyazuddin for valuable
discussion and also to Amjad Gilani for giving us the code [which he
developed during the review of Ref. \cite{Ali}] which is used to solve this
problem with some modifications. This work was supported by a grant from
Higher Education Commission of Pakistan.

{\LARGE Figure Captions}

\begin{enumerate}
\item[a]  Leading order contribution by operator $O_{7}$

\item[1]  Feynman diagrams contributing to the spectator corrections
involving the $O_{7}$ operator in the decay $B\rightarrow K_{1}\gamma $. The
curly (dashed) line here and in subsequent figures represents a gluon
(photon)

\item[2]  Feynman diagrams contributing to the spectator corrections
involving the $O_{8}$ operator in the decay $B\rightarrow K_{1}\gamma $. Row
a: Photon is emitted from the flavor changing quark line

Row b: Photon radiation off the spectator quark line

\item[3]  Feynman diagrams contributing to the spectator corrections
involving the $O_{2}$ operator in the decay $B\rightarrow K_{1}\gamma $. Row
a: Photon is emitted from the flavor changing quark line

Row b: Photon radiation off the spectator quark line

\item[4]  Feynman diagrams contributing to the spectator corrections in $%
B\rightarrow K_{1}\gamma $ decays involving the $O_{2}$ operator for the
case when both the photon and the virtual gluon are emitted from teh
internal (loop) quark line

\item[5]  Feynman diagrams contributing to the spectator corrections in $%
B\rightarrow K_{1}\gamma $ decays involving the $O_{2}$ operator for the
case when photon is emitted from the internal (loop) quark line in the $%
bs\gamma $ vertex.

\item[6]  The function $h_{0}(z)$ is plotted against the ratio $%
m_{f}^{2}/m_{b}^{2}$, where $m_{b}$ is the $b$-quark mass. The solid curve
is the real part of the function and the dashed curve is the imaginary part.

\item[7]  The function $h^{(K_{1})}(z,\mu _{\text{sp}})$ is plotted against
the ratio $m_{f}^{2}/m_{b}^{2}$ at the mass scale of the hard spectator
correction $\mu _{\text{sp}}=1.52$ GeV. The solid curve is the real part of
the function and the dashed curve is the imaginary part

\item[8]  The function $h_{1}(z)$ (left figure) and $h_{2}(z)$ (right
figure) are plotted against the ratio $m_{f}^{2}/m_{b}^{2}$, where $m_{b}$
is the $b$-quark mass. The solid curves are the real parts of the function
and the dashed curves are the imaginary parts.
\end{enumerate}

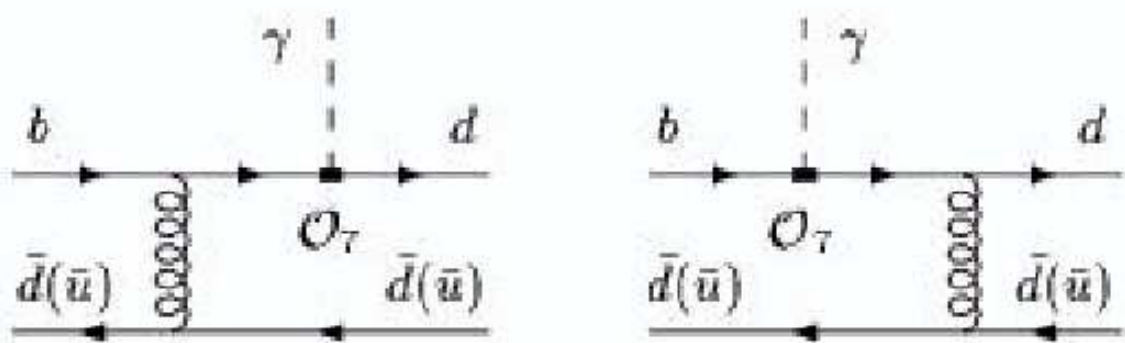
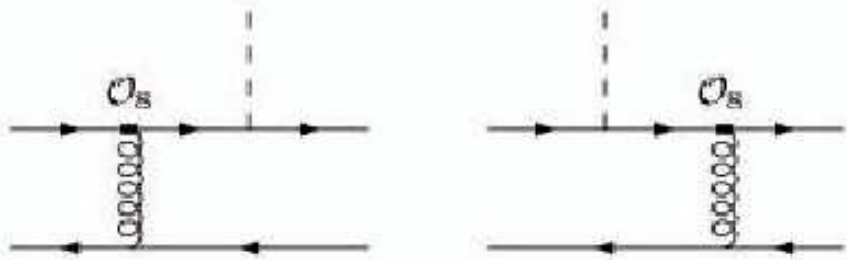
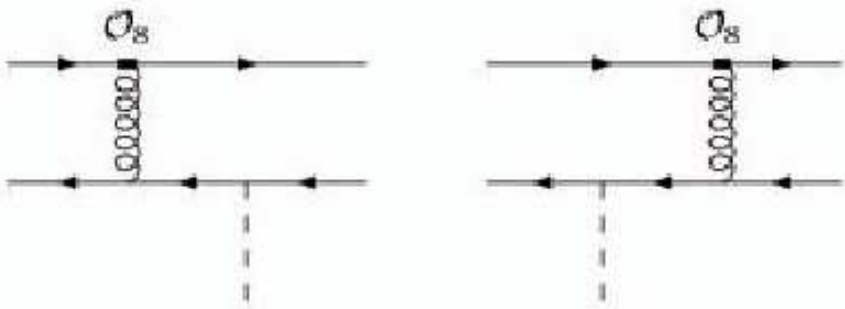
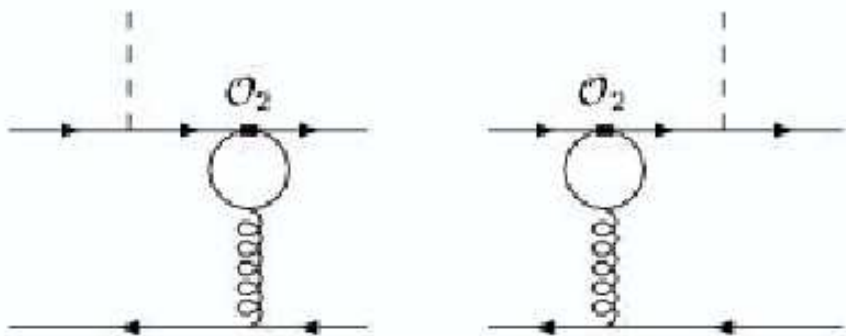
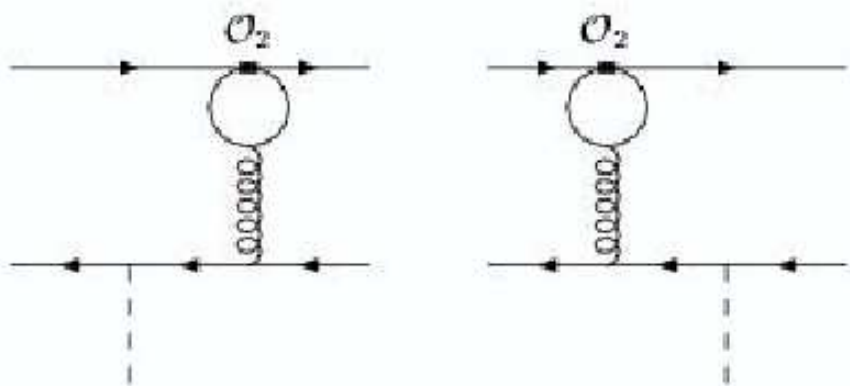
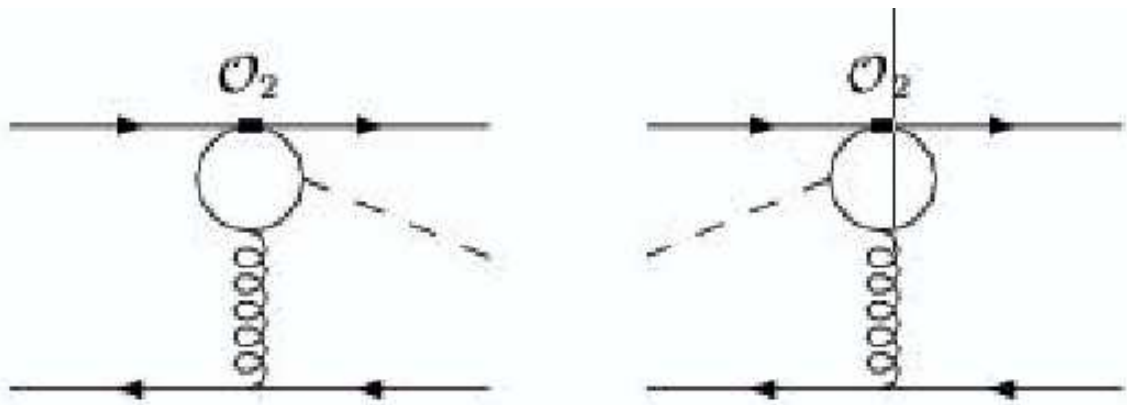
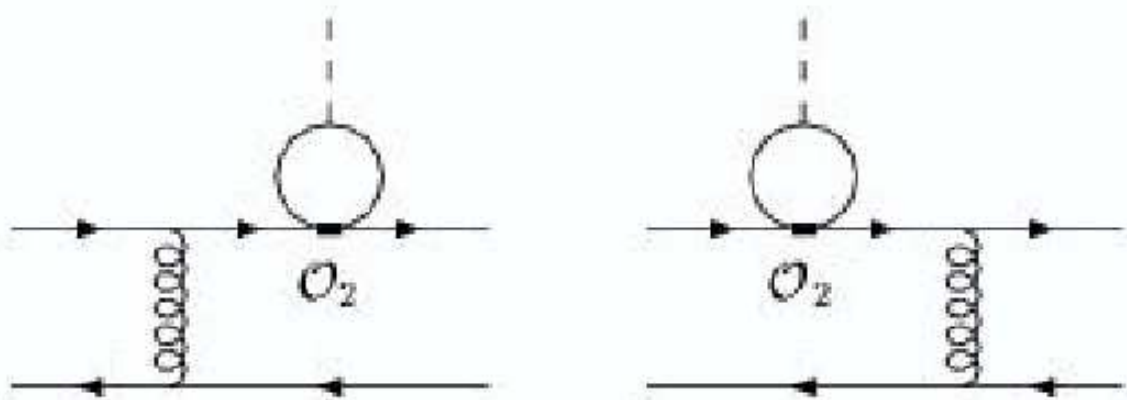
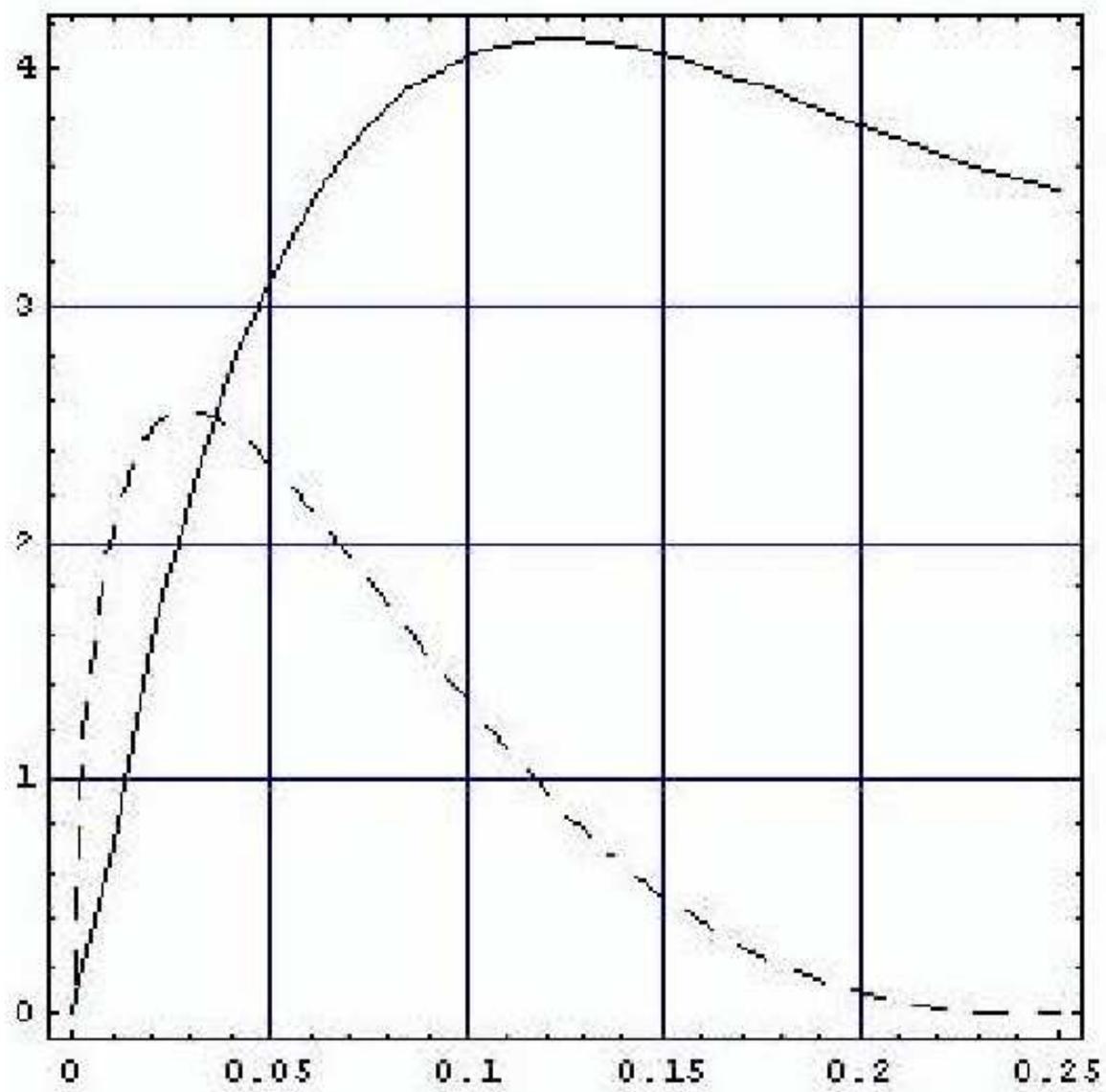
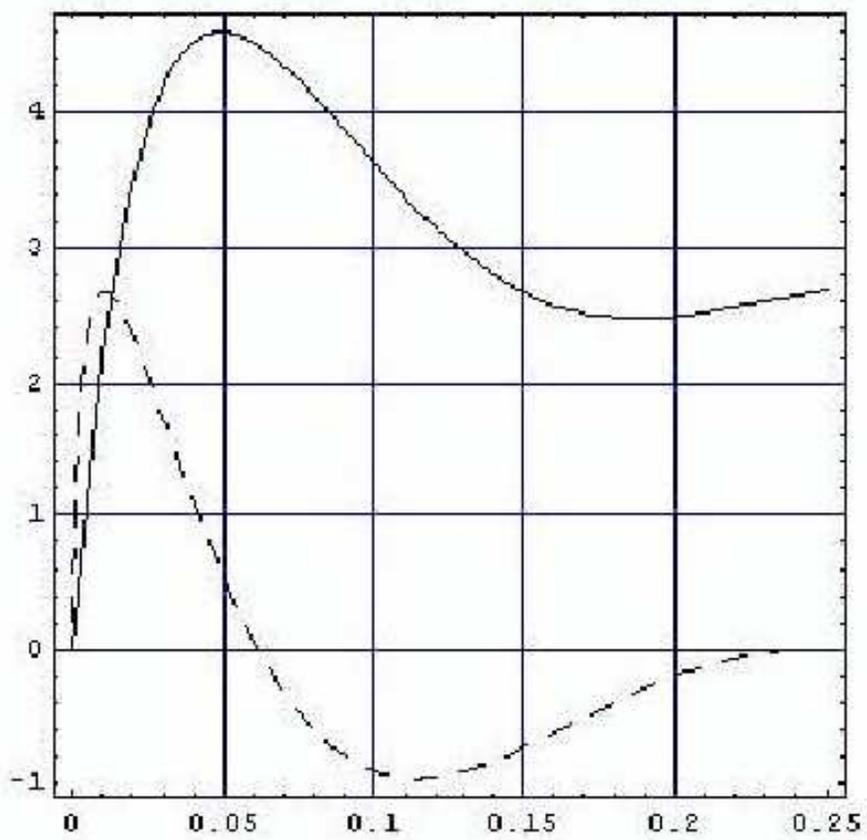
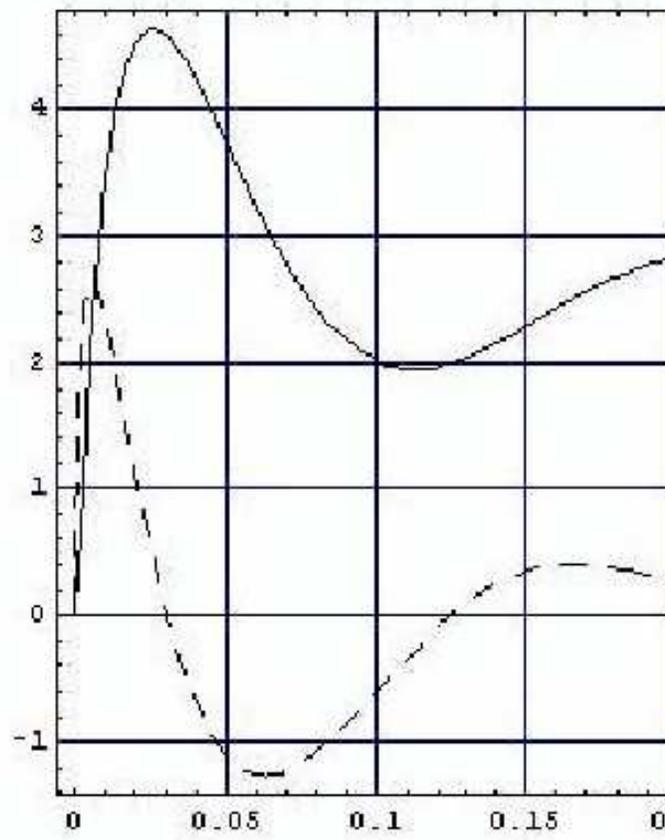
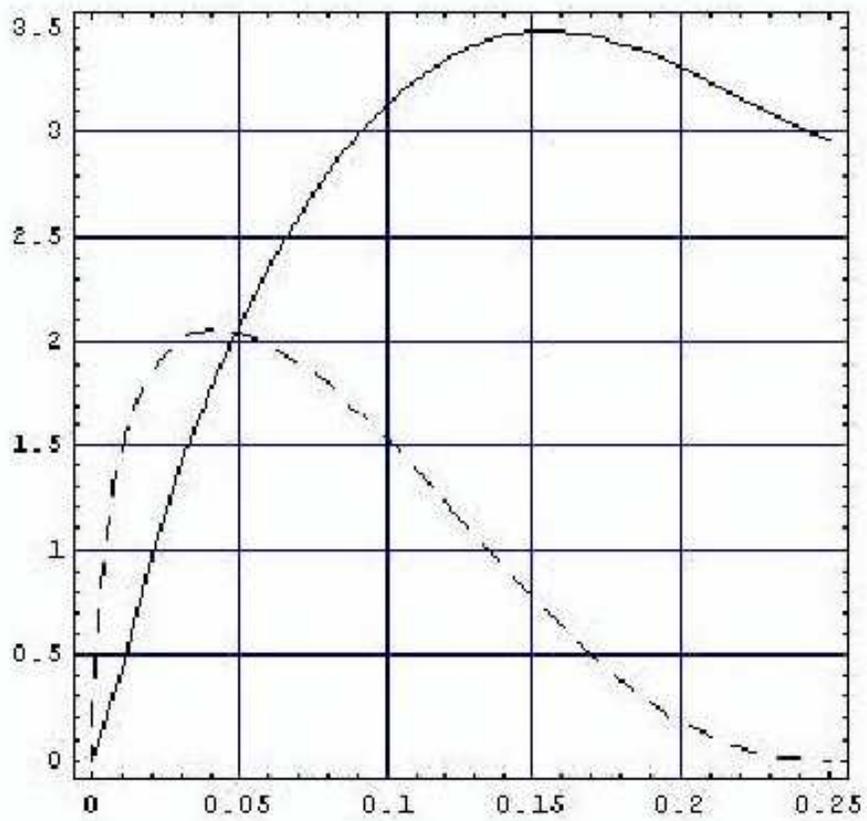
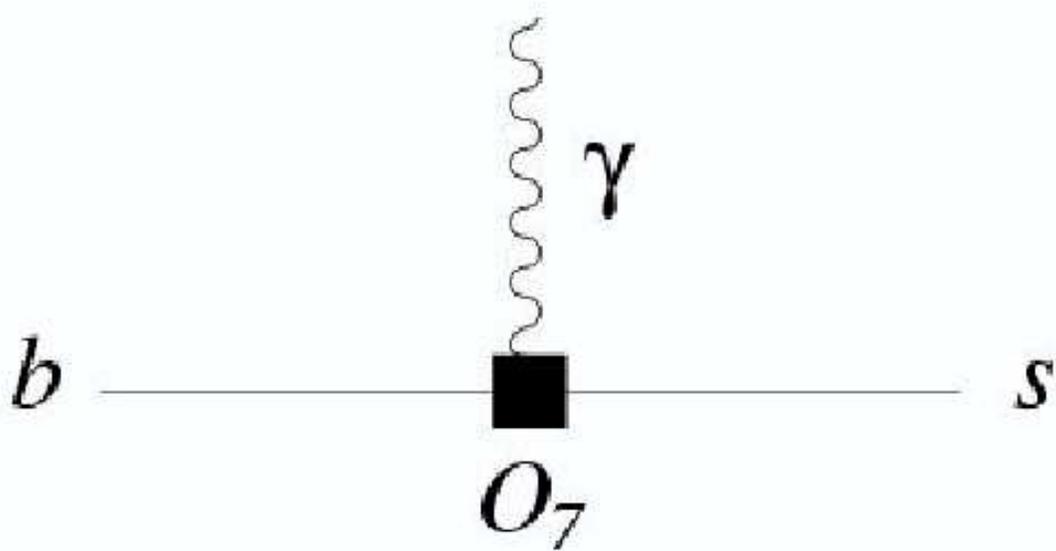
\end{document}